\documentclass[12pt, reqno, twoside]{amsart}
\usepackage{hyperref}

\usepackage{multicol}
\usepackage{amsmath, amsthm, amscd, amsfonts, amssymb, graphicx, color}
\usepackage{mathrsfs}
\theoremstyle{definition}

\theoremstyle{remark}

\numberwithin{equation}{section}

\begin{document}

\setcounter{page}{1}

\title[\hfilneg LAGJMA-2021/01\hfil Unilag Journal of Mathematics and Applications]{Two Pareto Optimum-based Heuristic Algorithms for Minimizing Tardiness and Late Jobs in the Single Machine Flowshop Problem}

\author[M. Gradwohl]{Matthew Gradwohl$^1$}
\author[G. Sewa]{Guidio Sewa$^2$}
\author[O. B. Oghojafor]{Oke Blessing Oghojafor$^3$}
\author[R. Wilouwou]{Richard Wilouwou$^4$}
\author[M. Adamu]{Muminu Adamu$^3$}
\author[C. Thron]{Christopher Thron$^*1$}

%\affil[1]{Department of Science and Mathematics, Texas A\&M University-Central Texas, Killeen TX USA}
%\affil[2]{Institut de Math{\'e}matiques et de Sciences Physiques, Porto Novo, Benin Republic}
%\affil[3]{Department of Statistics, University of Lagos, Lagos Nigeria}
%\affil[4]{University of South Brittany, Lorient France}

\address{Matthew Gradwohl \newline 
Department of Science and Mathematics, Texas A\&M University-Central Texas, Killeen TX USA}

\address{Guidio Sewa \newline 
Institut de Math{\'e}matiques et de Sciences Physiques, Porto Novo, Benin Republic.}

\address{Oke Blessing Oghojafor \newline 
Institut de Math{\'e}matiques et de Sciences Physiques, Porto Novo, Benin Republic.}

\address{Richard Wilouwou \newline 
University of South Brittany, Lorient France.}

\address{Muminu Adamu \newline 
Institut de Math{\'e}matiques et de Sciences Physiques, Porto Novo, Benin Republic.}

\address{Christopher Thron$^*$ \newline 
Department of Science and Mathematics, Texas A\&M University-Central Texas, Killeen TX USA}
\email{thron@tamuct.edu} 

\thanks{Submitted: April 1, 2021. Revised: Apri 10, 2021. Accepted: April 17, 2021.}  
\thanks{$^*$ Correspondence}

%%%%%%%%%%%%%%%%%%%%%%%%%%%%%%%%%%%%%%%%%%%%%%%%%%%%%%%%%
%%%%%%%%%%%%%  ABSTRACT %%%%%%% %%%%%%%%%%%%%%%%%%%%%%%%%
%%%%%%%%%%%%%%%%%%%%%%%%%%%%%%%%%%%%%%%%%%%%%%%%%%%%%%%%%

\begin{abstract}
Flowshop problems play a prominent role in operations research, and have considerable practical significance. The single-machine flowshop problem is of particular theoretical interest. Until now the problem of minimizing late jobs or job tardiness can only be solved exactly by computationally-intensive  methods such as dynamic programming or linear programming. In this paper we introduce, test, and optimize two new heuristic algorithms for mixed tardiness and late job minimization in single-machine flowshops. The two algorithms both build partial schedules iteratively. Both also retain Pareto optimal solutions at intermediate stages, to take into account both tardiness and late jobs within the partial schedule, as well as the effect of partial completion time on not-yet scheduled jobs. Both algorithms can potentially be applied to scenarios with hundreds of jobs, with execution times running from less than a second to a few minutes. Although they are slower than dispatch rule-based heuristics, the solutions obtained are far better. We also attempted a neural-network solution which performs poorly, and propose reasons why neural networks may not be a suitable approach.
\end{abstract}
\maketitle

%%%%%%%%%%%%%%%%%%%%%%%%%%%%%%%%%%%%%%%%%%%%%%%%%%%%%%%%%
%%%%%%%%%%%%%  SECTION - INTRODUCTION %%%%%%%%%%%%%%%%%%%
%%%%%%%%%%%%%%%%%%%%%%%%%%%%%%%%%%%%%%%%%%%%%%%%%%%%%%%%%
 
\section{Background  and motivation}

\subsection{The flowshop problem: an example of a scheduling problem}
In general, a \emph{scheduling problem} consists of organizing the execution of a sequence of jobs over time, taking into account time constraints (deadlines, sequence constraints) and constraints related to the availability of the required resources. This problem framework can be applied to a variety of practical situations, by modifying the interpretation given to jobs, machines and the necessary resources. Scheduling problems comprise a widely studied and complex combinatorial optimization problem. 

An important example of a scheduling problem is the \emph{flowshop problem}. In general, a \textit{flowshop} may be described as a system in which several ``jobs'' require processing on several sequential ``machines''. In different circumstances ``machines'' may refer to various types of processors (factory robots, line workers, editors, computer processors, etc.) that perform specified examinations or modifications of objects (automobiles, products, documents, computer programs, etc.), which are designated as ``jobs''.  In the flowshop problem, all jobs must pass through all the machines one by one in the same sequence.  A machine can only process one job at a time.  In order to be processed on machine $m$, the job must first pass through machines $1,\ldots,m-1$,  and it is not possible for two machines to work on the same job at the same time. In practice, flowshop problems occur in a wide variety of manufacturing industries including fabrics, chemicals, electronics, automotive, (\cite{bargaoui2016minimizing},\cite{oukil2021ranking},\cite{kurniawan2014optimizing}), iron and steel \cite{marichelvam2014solving}, food processing \cite{benthem2021solving}, ceramic tile \cite{chen2021population} , packaging and branding, pharmaceuticals, and paper manufacture \cite{ackermann2020analyzing} among others.

The flowshop problem is an example of a \textit{matching problem}: for each machine, each job must be matched with a position. The solution to a matching problem can be expressed mathematically as a \emph{permutation}. For example if we have three jobs, we may index them as $1,2,3$.  If job 2 is executed first, then job 3 and finally job 1, then this job ordering may be described using the permutation $(2,3,1)$.

\subsection{Common variants of the single-machine\\ flowshop problem}\label{sec:variants}

In different practical situations the flowshop managers may have different objectives. In some cases, the objective may be to minimize \emph{makespan}, defined as the total time required to complete all jobs.  Alternatively, and the manager may be primarily interested in meeting the customers' requested deadlines.  

 There is an extensive literature on solution methods for variants of the flowshop problems, for single or multiple machines. 
For the makespan minimization single-machine flowshop problem, an efficient exact  algorithm has been found. \cite{azad2021comparative} 
 A comprehensive review and evaluation of several heuristics and metaheuristics for the flowshop problem with total tardiness minimization may be found in \cite{vallada2008minimising}. Benchmarks for the permutation flowshop problem are presented in \cite{taillard1993benchmarks} and \cite{vallada2015new}.

In this paper, we consider situations where all jobs have due times, and each tardy job has a fixed penalty for being tardy, plus an additional penalty that is proportional to the amount of time that it is tardy. There is no bonus for jobs that are finished early.
For example if the  fixed penalty is 10 and the tardiness penalty rate is 5, then a job that is 3 days tardy has cost $10 + 3\cdot 5 = 25$.

\section{Methods}
\subsection{Mathematical specification of flowshop model}\label{sec:mathFormulation} %including information delay.

\subsubsection{Constants and variables}

Our flowshop model can be expressed mathematically in terms of the following constants and variables.

\noindent \textbf{\textit{Constants:  }}
All constants listed here are positive real numbers.
\begin{itemize}
    
\item $a_1,\dots a_N$: Arrival times for jobs at the flow shop.  Jobs are listed in order of arrival, so the $n$'th job to arrive has index $n$; 

\item $d_1,\dots d_N$: Due times, where $d_n$ is the due time for the job that has arrival time $a_n$

\item $x_{1},\dots x_{N}$: Processing times, where $x_{n}$ is the processing time of the job with index $n$.

%\item $z_{11}\ \dots .z_{NM\ }$: Waiting times, where $z_{nm}$ is the waiting time of the job with index $n$ before machine $m$.

\item \noindent $p$:  fixed penalty (cost) if a job is tardy;  

\item  $q$:  Tardiness penalty coefficient if a job is tardy. For example, if job $n$ arrives at time $d_n+t$, then the tardiness penalty for that job is $qt$ (this is in addition to the fixed penalty $p$).

\end{itemize}

\noindent \textbf{\textit{Variables: }}

\begin{itemize}
\item
Tardiness: $w_k\in \ {\mathbb{R}}^+$  $w_k$ is the tardiness of the $k$'th job to finish. Note that this is not the same as the job with index $n$.  The tardiness is always nonnegative ($N$ real variables).
\item Matching variables: $u_{nk}\in \left\{0,1\right\},$ where $1\le n,k\le N$.  $u_{nk}=1$ if job $n$ is the $k$'th job scheduled. Otherwise, $u_{nk}=0$ (${N}^{2}$ binary variables).
\item Starting times: $s_{nk} \in \ {\mathbb{R}}^+ $: Starting times, where $1\le n,k\le N$. If $u_{nk}$ is 0, then $s_{nk}$ is also 0. Otherwise, $s_{nk}$ is the starting time of job $n$  (note that job $n$ is the $k$'th job to execute (${N}^{2}$ real variables). Under this definition, $\sum_k{s_{nk}}$ will be the starting time of job $n$,  and $\sum_n{s_{nk}}$ is the starting time of the $k$'th job scheduled.
\item  Tardy indicators: $v_k\in \{0,1\}$; $v_k=1$ if the $k$'th job to complete  is tardy, and 0 otherwise ($N$ binary variables).
\end{itemize}

\subsubsection{Objective function} \label{sec:objFunction}

In this paper, we will consider the problem of minimizing a penalty  that depends both on the number of tardy jobs and on tardiness, as described in Section~\ref{sec:variants}.  
In the mathematical specification, this corresponds to the following objective function:

\begin{equation} \sum^N_{k=1}{q\cdot w_k+}\sum^N_{k=1}{{p\cdot v}_k} 
\end{equation}

\subsubsection{Constraints}

\begin{itemize}

\item
\textit{Matching conditions} ($2N$ equality constraints):  
\begin{equation}
\sum_n{u_{nk}=1\ \forall \ k;\ \ \sum_k{u_{nk}=1\ \forall \ n}}
\end{equation}

\item 
\textit{Lateness lower bound for $k$th job scheduled} ($N$ inequality constraints):  
\begin{equation}
-w_k +\sum_n{\left(s_{nk}+u_{nk}\left(x_{n}-d_n\right)\right)\ }\le 0,\ \ \ \ \ 1\le k\le N.
\end{equation}

\item
 \textit{Positivity of tardiness for job in position $k$} ($N$ inequality constraints):  
 \begin{equation}
{-w}_k\le 0.  
\end{equation}
This constraint and the previous constraint ensure the correct value of tardiness.

\item 
\textit{Tardy binary indicators} ($N$ inequality constraints):
\begin{equation}
    w_k{-C_{big}}\cdot v_k\le 0\ \forall k,
\end{equation}\ 
where $C_{big}$ is a very large number. This constraint guarantees that $v_k=1$ whenever $w_k>0$, and otherwise $v_k=0$.

\item 
\textit{Starting time follows arrival} ($N$ constraints):
\begin{equation}
-\sum_k{s_{nk}\ }\le -a_n.    
\end{equation}

\item
\textit{Starting time constraint for consecutive jobs} ($N-1$ constraints):
\begin{equation}
  \sum_n{(s_{nk}-s_{n(k+1)})}+\ \sum_n{u_{nk}x_{n}}\le 0,\ \ \ \ \ k<N.    
\end{equation}
(A job cannot start until the previous job is finished.) 

\item
\textit{Nonnegative starting times}($N^2$ constraints):
\begin{equation}
-s_{nk}\le 0.    
\end{equation}

\item
\textit{Starting times agree with matching variables} ($N^2$ constraints): 
\begin{equation}
s_{nk}-C_{big}\cdot u_{nk}\le 0,    
\end{equation}
where $C_{big}$ is a very large number.  (This guarantees that $s_{nk}$ is only positive when $u_{nk}=1$.) 

Note that constraint (4) is actually implied by the requirement that $w_k\in \ {\mathbb{R}}^+$--we have included it in the constraint list for clarity.

\end{itemize}

\subsection{ Algorithms for solution and approximate solution}
The mathematical problem described in Section~\ref{sec:mathFormulation} can be solved exactly using mixed-integer linear programming (MILP), since both the objective function and all of the constraints described in Section \ref{sec:mathFormulation} are linear.  However, the solution is computationally prohibitive  except for very small systems.  For larger systems, we must rely on  approximate algorithms, which are typically either heuristic or stochastic.  A newer alternative to these conventional methods is neural network.

\subsubsection{Existing heuristic scheduling algorithms}\label{sec:ExistingHeuristics}

The simplest and fastest classic heuristics use dispatch rules to determine the job execution order. 
According to \cite{pinedo1992scheduling}, some dispatch rules that have been utilized include Earliest Due Time (EDD) (i.e., the job with the soonest due time runs first),  Shortest Processing Time (SPT) and Longest Processing Time (LPT) (i.e.  the fastest to process or slowest to process is run first, respectively). \cite{nahmias2016production} identifies an additional dispatch heuristic Critical Ratio, which takes the amount of time left to process each job before its due date and divides by the processing to give each job a priority index, with higher priority given to the job with the lowest priority index.
%Least Slack Time (LST) LST prioritizes the jobs based on slack time (the amount of time the job is available to process minus the time it will take to process; $d_n-a_n-x_n$), favoring the jobs with the least amount of slack time.

Iterative heuristics add some number of jobs at a time and evaluate possible orders before moving on, with the order to be tried  determined based on initial criteria, possibly using one of the above algorithms. Existing methods include the Palmer heuristic which calculates an index for each job and uses that to determine the initial order \cite{nugraheni2021combination}, and the NEH heuristic, which prioritizes jobs with longer executions over possibly multiple machines \cite{sauvey2020}. Until now it appears that the Palmer heuristic has not been adapted to tardiness minmization. On the other hand, there is a modification of NEH (called "NEHedd") that can be applied to tardiness minimization problems \cite{kim1993heuristics}. Like NEH, NEHedd proceeds by iteratively constructing longer and longer job sequences, such that the final constructed sequence is the estimated job order.  During the iterative process prescribed by NEHedd, it is often the case that more than one partial sequence meets the critierion for selection, i.e. there are "ties" between equally qualified partial sequences.  Various mechanisms can be used to resolve these "ties" \cite{fernandez2015neh}.   However, previous tie-breaking criteria  do not make use of Pareto optimality to take into account the multi-criteria nature of the intermediate problems.

\subsubsection{Neural network solution for single-machine problem}\label{sec:NNreplaceMILP}

\noindent \emph{Background}
\medskip

Neural networks (NN) have the potential of obtaining near-optimal solutions very quickly. 
The training process for a NN may be time-consuming, but once trained the NN can execute extremely rapidly, even for sophisticated, deep neural networks with millions of parameters.

We construct a NN to solve the single-machine flowshop problem as follows. 
%Note that it is not necessary to include information delay as a factor, because with a single machine there is no waiting between machines. Any information delay from the previous machine simply changes the arrival time to the machine in question.
\medskip

\emph{Neural network structure}

\begin{figure}[ht]
    \centering
    \includegraphics[width=2.1in]{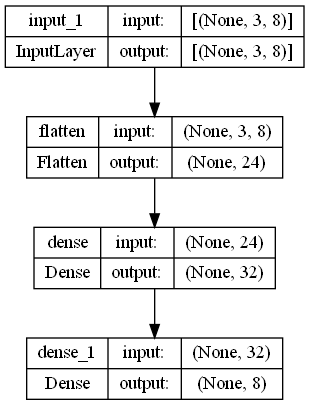}
    \caption{Neural network layer structure for a single-machine system with 8  jobs. Numbers in the right-hand boxes indicate the sizes of input and output arrays for each layer.}
    \label{fig:model_plot}
\end{figure}

The NN used is a shallow network consisting of 3 layers: an input layer, a hidden layer, and an output layer. Figure~\ref{fig:model_plot} displays the output of tensorflow's \texttt{summary} command showing the structure of the neural network, indicating the layers, connections, and input and output sizes.
A more complete description of the network is as follows.
\begin{itemize}
\item \emph{Input layer:}  
The inputs are arranged as  a $3 \times N$ matrix. The three rows of the matrix contain availability times,  processing times, and  due times respectively, with  column $j$ representing the job with index $j$. 
%Each row of the input matrix is exactly as generated, while the labels are divided by $N$ so that the outputs can be compared to the labels.

\item \emph{Hidden layer:}
The hidden layer is fully connected. The number of nodes and the number of hidden layers were selected based on performance for $N=8$ jobs.

\item \emph{Output layer:}
This layer is fully-connected with $N$ nodes, and a sigmoid activation function that produces values between 0 and 1. The estimated schedule is obtained by ordering the outputs.. 
For example, if the NN produces the output $[0.3, 0.4, 0.2]$ then the job order should be the third job (lowest value) followed by the first job  and lastly the second job (highest value). In the python code, this ordering is recovered using the command \texttt{argsort}.
\end{itemize}

\subsubsection{Pareto-based heuristics for single-machine problem}\label{sec:algDescriptions}

In this paper we propose two novel heuristic methods, which we designate as "iterative insertion" and "iterative selection". 

The iterative insertion algorithm is similar to the NEH algorithm that is used for makespan minimization. Modifications must be made to accommodate special difficulties posed by the tardiness-tardy time minimization criteria. In particular, since multiple criteria are used to determine ``good'' solutions at intermediate stages, we use Pareto optimality as a selection criterion to determine multiple candidate solutions.  

On the other hand, we did not see any algorithm in the literature that is comparable to iterative selection.  Iterative selection also relies on Pareto optimality to obtain a set of candidate solutions at each intermediate stage.  

In the following subsections, we provide detailed descriptions of the iterative insertion and iterative selection algorithms.
\medskip

\noindent \textbf{\emph{Iterative insertion}}
\medskip

 Iterative insertion proceeds as follows. First,  a preliminary ordered list of jobs is created, possibly using one of the simple heuristics mentioned in Section~\ref{sec:ExistingHeuristics}.  During the algorithm, jobs are selected from this preliminary list one by one.
 The algorithm creates a series of lists of ordered lists as follows.  
\begin{itemize}
\item
The first list of ordered lists is a list with a single element containing the first preliminary job. 
\item
The second list of ordered lists is formed as follows. Using the single list in the first list of lists (which at this stage contains a single element), insert the next preliminary job at each possible point in the list (at this point, there are two possible insertion points), to create two different lists of two elements. For each of these 2-element lists, evaluate the tardiness penalty and the finish time associated with these job orderings, and retain all of the Pareto-optimal orderings.
\item
Each subsequent stage of the algorithm resembles the second stage.  For stage $m$, the $m$'th element of the preliminary list is selected, and inserted at each possible point in all lists in the list of lists that was created at the $m-1$'th stage. For each of the resulting lists, evaluate the tardiness penalty and the finish time associated with the given job ordering, and retain all Pareto-optimal orderings.
\item
At the last stage,  retain only the lists with minimal tardiness penalty.
\end{itemize}

We may give a specific example of the algorithm as follows.  Suppose the preliminary order of jobs is $[4,3,2,1]$. At the first stage, the list of lists is $[[4]]$.  At the second stage, the lists $[4,3]$ and  $[3,4]$ are evaluated for tardiness and finish time.  Suppose that both tardinessses are 0, and the finish times are equal. Then both ordered lists are retained: so at the end of the second stage, the list of ordered lists is $[[4,3],[3,4]]$.  At the third stage, job 2 is inserted at each point within each list of ordered lists from the second stage. This produces a list of 6 ordered lists:  $[[2,4,3],[4,2,3],[4,3,2],[2,3,4],[3,2,4],[3,4,2]]$. Of these, select those job orderings which are Pareto optimal with respect to tardiness and finish time.  We may suppose for example that this reduces the list to 3 ordered lists:   $[[2,4,3],[2,3,4],[3,2,4]]$. Finally, we insert job 1 at each position in each of these lists, giving rise to 12 ordered lists. From these 12 ordered lists, we select one that has minimum tardiness, since each minimum-tardiness solution is equally optimal.

In the example above,  we note that an exhaustive evaluation of all possible job orderings would require $4! = 24$ evaluations for 4 jobs, considerably more than the 12 we ended up evaluating.  The savings for larger numbers of jobs depends on the number of ordered lists retained from stage to stage.  If for example this number is bounded by $B$, then the number of evaluations at stage $m$ is $(m+1)B$, so that the total number of evaluations grows quadratically with the number of jobs (rather than exponentially).

In the above algorithm, the starting list of ordered lists contains a single element, which consists of a single job.  we may define a variant of the algorithm as follows.  Instead of starting with a single-element list, take the first $J$ jobs in the preliminary list, and use as starting list the $J!$ permutations of the first $J$ jobs. From this list, retain the Pareto optimal orderings, and then continue the algorithm as described above. $J$ must not be chosen too large:  for example, $J=5$ means that the starting list has 120 permutations,  

When calculating the penalty and finish time for the different jobs with insertion, it is possible to reduce computation time by intelligently choosing the order of calculation. For example, if job 3 is to be inserted in the list $[2,6,4,5,1]$, then one may first calculate the penalty and finish time when the new job 3 is inserted  in the last place, i.e. $[2,6,4,5,1,3]$.  Next, insert the new job 3 in the second to last place, i.e. $[2,6,4,5,3,1]$. Then one may reuse the calculated finish times for $2,6,4,5$, because these jobs' starting and finishing times remain unchanged. If the calculation is done this way, the complexity is reduced by a factor of 2.

Similarly, if the variant where the starting list consists of $J$ jobs the finish times and tardinesses for the $J!$ orderings may be calculated more efficiently by ordering the calculations properly.  For example, consider the case where one wants to calculate the penalties and finish times for all permutations of $[1,2,3,4]$.  We may first calculate and store the penalties and finish times for each of the single-element ordered lists, namely $[1]$, $[2]$, $[3]$, and $[4]$. Then for example we may use the result for $[1]$ to calculate the finish times and penalties for the 2-element ordered lists that start with 1, namely  $[1,2]$, $[1,3]$, $[1,4]$.  Similarly, we may use $[2]$ to calculate the finish times and penalties for the 2-element ordered lists that start with 2, namely $[2,1]$, $[2,3]$, $[2,4]$.  We may do the same with $[3]$ and $[4]$.  Then we may use the results from $[1,2]$ to calculate results for $[1,2,3]$ and $[1,2,4]$: and so on.  This method will reduce calculation by 1/2.

Two modifications were introduced to limit the complexity of the algorithm:
\begin{itemize}
\item
  It is possible that after several iterations, the set of Pareto optimal candidates may become unmanageably  large. We counteract this effect by limiting the number of Pareto optimal solutions retained.  This is done by retaining a representative subset of permutations from the Pareto optimal list for each iteration. We thus introduce a tuneable parameter $P$ correponding to the maximum number of permutations retained.
In order to maintain a variety among the sequences that are retained,  the Pareto optimal candidates are sorted by completion time, and every $\lfloor L/P \rfloor$'th sequence is retained, where $L$ is the number of candidates.
\item
One can also limit the insertion slots to a set number $S$, so that new jobs are only inserted into the last $S$ possible positions in the previously-generated job sequences. In especially large numbers of jobs generated with a similar distribution to the training data, sorting by relative due time will put jobs that can be run later closer to the end, and they will not need to be tried at the beginning, as that would just push the other jobs to be tardy. Rather than trying those jobs at the beginning of the list, time is saved by just inserting in the specified number of insertion slots from the end.
\end{itemize}
The MILP formulation requires $O(N^2)$ variables and constraints, making
exact solutions impractical for $N > 10$ (e.g., $N = 50$ yields $5,100$ variables). In contrast,
the insertion heuristic reduces this to $O(P \cdot S \cdot N)$. If both of these modifications are implemented in the insertion algorithm, then at each iteration the number of sequences to be calculated is $P \cdot S$, and the length of the sequence is $S$ (since it is not necessary to recalculate the job sequence before the first insertion point). It follows that the algorithm's execution time will be linear in the number of jobs.
\medskip

\noindent \textbf{\emph{Iterative selection}}
\medskip

Instead of successively building up schedules by inserting new jobs, an alternative approach is based on appending new jobs to candidate partial schedules sequentially by a trial-and-retain procedure.  For example, suppose it is desired to schedule 7 jobs $[1,2,3,4,5,6,7]$, and suppose a preliminary ordering (determined by a simple heuristic, such as greedy) has the jobs  in order $[2,1,5,6,3,7,4]$.  Then for the first $W=4$ jobs, (namely $[2,1,5,6]$ in this case) we may first exhaustively evaluate the  penalties and finish times for all $W!=24$ possible orderings of $[2,1,5,6]$.  From this list, we may retain the Pareto optimal orderings (with respect to penalty and finish times), and choose the initial job in each of these Pareto optimal orderings.  This will give us a list of single-element orderings. Suppose for example, that 2 and 5 are the initial elements of Pareto optimal orderings of $[2,1,5,6]$.  We then evaluate finish times and penalties for all orderings of the first five jobs in the preliminary list (namely 2,1,5,6,3) which begin with either 2 or 5.   This requires evaluations for $2W!$ orderings.  From these, we retain the Pareto optimal orderings, and select the first 2 elements for all of these Pareto optimal orderings.  With reference to our example, suppose $[2,6,5,1,3]$, $[2,1,5,6,3]$ and $[5,2,6,1,3]$ are Pareto optimal. Then we will retain the 2-element ordered lists $[2,6]$, $[2,1]$, and $[5,2]$ as candidates for the first two jobs.  At the next stage, we evaluate penalties and finish times for all ordered sequences involving the jobs $[2,1,5,6,3,7]$ that begin with either $[2,6]$, $[2,1]$, or $[5,2]$: this involves calculating $3W!$ penalties and finish times. As before, we retain only the first three elements of all Pareto optimal ordered sequences.  Suppose for example that all Pareto optimal sequences at this stage begin with $[2,1,6]$.  Then at the next stage, we only consider ordered sequences of length 7 that  begin with $[2,1,6]$. 

As with interative insertion, it is possible that the number of Pareto optimal candidates will grow too large.  In this case we may use the same strategy as before: namely order the Pareto optimal candidates by completion time, and choose $K$ regularly-spaced job sequences from among the candidates.  It follows that the iterative selection algorithm also has two tuneable parameters, namely the permutation window and the number of retained sequences, which we denote as $W$ and $K$, respectively.
 
\subsection{Training, testing, and evaluation of heuristic methods}

\subsubsection{Construction of scheduling scenarios for training and testing}
In order to test the heuristic insertion and train the neural network described in Section~\ref{sec:NNreplaceMILP}, %algorithms described in the above sections, (as well as to train the neural network described in Section~\ref{sec:NNreplaceMILP}), 
it is necessary to generate a set of labeled instances for scheduling. The instances are generated with randomized arrival, execution, delay, and due times, as summarized in Table~\ref{tab:1}, which follows the notation of Section~\ref{sec:mathFormulation}.
% follows (following the notation in Section~\ref{sec:mathFormulation}) 
% \begin{itemize}
% \item
% The job arrivals $a_n, n=1\ldots N$ are generated as a Poisson process whose interarrival times are independent exponential random variables with mean $\mu_A$; 
% \item
% The execution times $x_{n,m}, n=1,\ldots N, m = 1 \ldots M$ are  exponentially distributed with mean $\mu_X$;
% \item
% The waiting times $z_{n,m}, n=1,\ldots N, m = 1 \ldots M$ are  exponentially distributed with mean $\mu_Z$;
% \item
 The due times $d_n, n=1,\ldots N$ are generated as follows:

\begin{equation}
    d_n = a_n + z_{n}+ x_{n} + D'_n,
\end{equation}
where the random variable $D'_n$ is exponentially distributed with mean $\mu_{D'}$.  The role of $D'_n$ is to allow some margin for each job, so that job $n$ can experience a net delay of $D'_n$ or less and still be completed on time. The parameter values used are summarized in Table~\ref{tab:1}. 
All algorithms are implemented using Python 3.10, with the MILP solver from Scipy 1.12. The Spyder integrated development environment (version 5.5.1) was used.

\begin{table}
    \centering
    \begin{tabular}{|l|c|c|}
    \hline\\
        Variable name & Distribution & Parameter \\
        \hline\\
        Arrival times & $a_1\sim \text{Exp}(\mu_{A});~~  a_n\sim a_{n-1} + \text{Exp}(\mu_{A})$ & $\mu_A = 5$ \\
        \hline\\
        Execution times & $x_{n} \sim \text{Exp}(\mu_X)$ & $\mu_X = 5$\\
        \hline\\
        Information delays & $z_{n} \sim \text{Exp}(\mu_Z)$ & $\mu_Z = 5$\\
        \hline\\
        Delay margins & $d'_n \sim \text{Exp}(\mu_{D'})$ & $\mu_{D'} = 10$\\
        \hline\\
        Due times & $a_n + x_{n} +  z_{n}  + d'_n$ & -- \\
        \hline
    \end{tabular}
    \caption{Random variables and parameters for random generation of arrival, execution, delay times, delay margin, and due times. The index $n$ labels jobs and runs from $1$ to $N$.}\label{tab:1}
\end{table}

% \begin{table}
% \begin{center}
%     \begin{tabular}{ | m{35em} | }
%         \hline\\
%         Arrival times: $a_1\sim \text{Exp}(\mu_{A});~~  a_n\sim a_{n-1} + \text{Exp}(\mu_{A})~~n=2\ldots N$, where $\mu_A = 5$.\\ 
%         \hline\\ 
%         Execution times:  $x_{n,m} \sim \text{Exp}(\mu_X),~ n = 1,\ldots N, m=1\ldots M$,where $\mu_X = 5$\\
%         \hline\\
%         Information delay  before machines: $z_{n,m} \sim \text{Exp}(\mu_Z)~ n= 1\ldots N, m=1\ldots M$, where $\mu_Z = 5$.\\
%         \hline\\
%         Delay margin: $D'_n \sim \text{Exp}(\mu_{D}),~n=1\ldots  N$ \\
%         \hline\\
%         Due times:
%         $d_n:= a_n + \sum_{m=1^M}x_{n,m} +  \sum_{m=1^M}z_{n,m}  + D'_n, n=0\ldots N. $\\
%         \hline
%     \end{tabular}
%     \caption{Random variables and parameters for random generation of arrival, execution, delay times, and due times}\label{tab:1} 
% \end{center}
% \end{table}
%\multirow{2}{5cm}{

\subsubsection{Training the neural network}

The exact and heuristic single-machine algorithms do not need training.  Only the neural network model needs training, and in this subsection we describe the training procedure.

$12,000$ single-machine cases with $6$ jobs ($N=6$) were generated according to the distributions in Table~\ref{tab:1}. The optimal job orders for these single-machine cases were computed using the exact MILP algorithm. These optimal orders were used to create labels for each case as follows. Let $\sigma$ denote the permutation corresponding to the correct ordering for a particular case, i.e. $\sigma[j]=k$ if the $j$'th job scheduled is job $k$. Then the label for that case is a vector $L$ of length $N$ with entries in $[0,1]$ given by:
\begin{equation}
    L[j] := \frac{k}{J-1},
\end{equation}
so that the order of the entries in $L$ matches the order of the jobs scheduled. 
The metric used to evaluate the  NN output is mean squared difference between the label and the output.

To improve the training performance of the NN, first the starting structure is adjusted. In initial testing, we tried layers with between $4N$ and $8N$ nodes, and with linear, softmin, softmax, softplus, softsign, hyperbolic tangent, SELU, elu, ReLU, and sigmoid activation functions. These functions were selected because of their common use, although other activation functions exist. More limited testing on the number of hidden layers suggested that one hidden layer with $4N$ nodes and a linear activation has the best performance using the objective function listed in \ref{sec:objFunction}. Parameter optimization methods tried include adadelta, adagrad, adam, adamw, adamax, RMSProp, and Stochactic Gradient Descent (SGD). The batch size was not optimized, and was set at 1. The mean squared error (MSE) loss function was used to gauge performance. The number of epochs was adjusted based on the loss curves, which leads to an epoch value of $250$ where the curves begin to plateau. 

\begin{figure}[ht]
    \centering
    \includegraphics[width=3in]{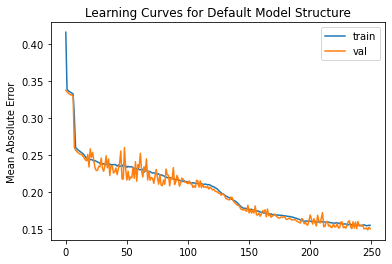}
    \caption{Learning loss curves for NN. The close match between training and validation curves show that overfitting has not taken place.}
    \label{fig:default_model_learning}
\end{figure}

During training we plot the loss curves to assist in evaluating the number of epochs. The loss curve is shown in  Figure~\ref{fig:default_model_learning}, which  shows the flattening of the loss curve at 250 epochs.

\subsection{Performance Comparisons}

In order to evaluate the relative performance of the different algorithms, a number of tests were performed. All test scenarios were generated using the distributions and parameters described in Table~\ref{tab:1}.
All comparisons and training were run on a workstation with 32.0 GB of RAM and an Intel i7-1360P CPU.

First we compared numerical algorithm performance to the exact MILP solution by applying MILP, NN, and insertion to  the same 2000 scenarios with 6 and 8 jobs, and comparing the objective functions obtained by the three methods. The number of jobs was limited because of the high runtimes of MILP. 

Second,  we compared the selection and insertion heuristic algorithms by applying both  to the same $100$ instances of $50$ jobs. Both algorithms were run with different values for the algorithm parameters: maximum number of kept permutations and maximum number of insertion slots for the insertion algorithm; maximum number of kept permutations and selection window for selection.  Several graphs comparing performance and execution times for the different algorithm variants were generated.

\section{Results}
\subsection{Comparison of numerical algorithms with exact MILP solution for small scenarios}
\begin{figure}[ht]
	\centering
	\includegraphics[width=0.6\linewidth]{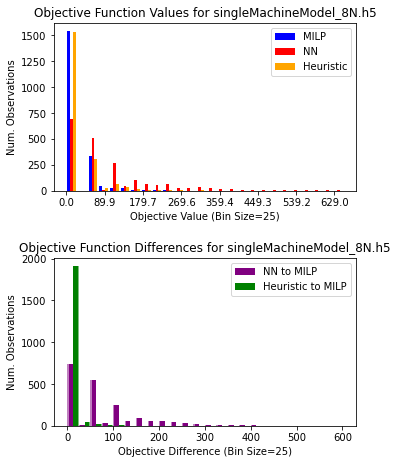}
	\vspace{-3mm}
	\caption{Distributions of objective function differences (NN minus MILP and insertion heuristic minus MILP), applied to 2000 instances of 8 jobs with distributions according to Table~\ref{tab:1}.}
	\label{fig:obj_compar}
\end{figure}

Figure~\ref{fig:obj_compar} shows the distributions of objective functions obtained using the NN, MILP, and insertion heuristic for 2000 instances of 8-job scenarios with parameters given by Table~\ref{tab:1}. For the insertion heuristic since the number of jobs was limited all permutations were kept, and all insertions were included. Clearly the insertion algorithm performance is almost identical to that of the exact MILP, while the NN is far less optimal. 

\begin{figure}[ht]
	\centering	\includegraphics[width=0.8\linewidth]{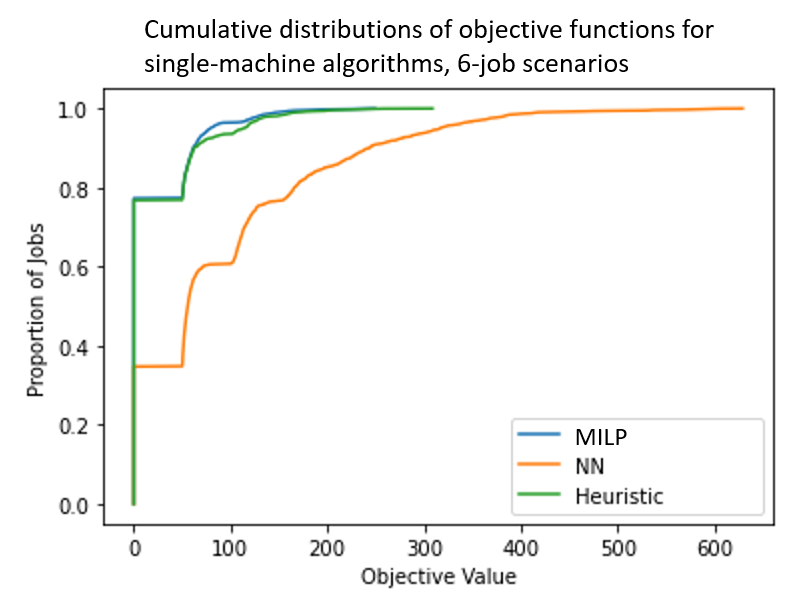}
	\vspace{-3mm}
	\caption{Cumulative distributions of objective functions for MILP, NN and insertion heuristic single-machine algorithms, applied to 2000 instances of 8 jobs with distributions according to Table~\ref{tab:1}.}
	\label{fig:cum_compar}
\end{figure}

Figure~\ref{fig:cum_compar} displays the same information  as Figure~\ref{fig:obj_compar} for the 6-job case, but in the form of cumulative distributions. Here we can see more clearly the close agreement between the exact MILP solution and the heuristic solution. We conclude that at least for few enough jobs, the insertion heuristic is usually giving us an  overall best solution.

\subsection{Performance comparisons between insertion and selection algorithms for larger cases}

\subsubsection{Objective function comparisons}

\begin{figure}[ht]
	\centering	\includegraphics[width=1.0
    \linewidth]{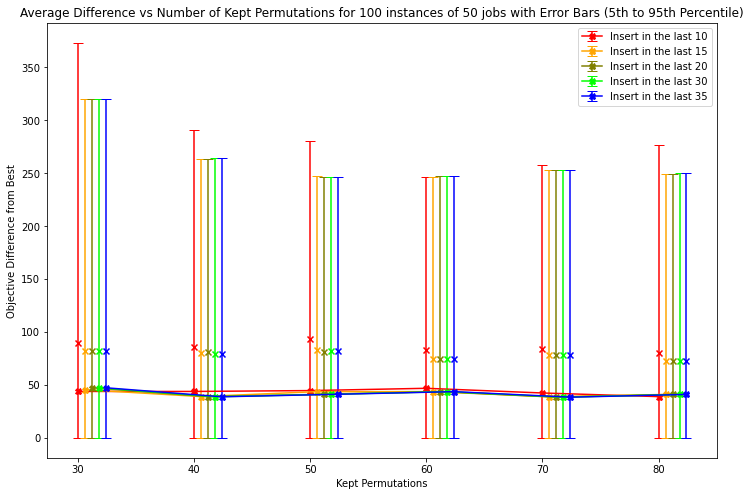}
	\vspace{-3mm}
	\caption{Performance comparison for 30 insertion algorithm variants using different parameter values (number of kept permutations $P$ and number of insertion slots $S$), for 100 50-job scenarios. The number of insertion slots ($S$) is listed on the $x$ axis, while different values of $P$ correspond to the different series shown in the legend. The $y$ axis shows the difference between the objective function obtained by the algorithm and the lowest overall objective function obtained from all 30 insertion algorithm variant shown in this figure  and the 18 different selection algorithm variants shown in Figure~\ref{fig:AvgObjDif_NKEEP_Select} for each 50-job scenario.  Error bars show the 5th and 95th percentiles for each algorithm variant. Medians for each algorithm variant are plotted with circles and are joined with lines, while means are plotted with $\times$'s. Variants with  $S \ge 10$ and $P \ge 60$ have similar performance statistics.}
	\label{fig:AvgObjDif_NKEEP_Insert}
\end{figure}

\begin{figure}[ht]
	\centering	\includegraphics[width=1.0
    \linewidth]{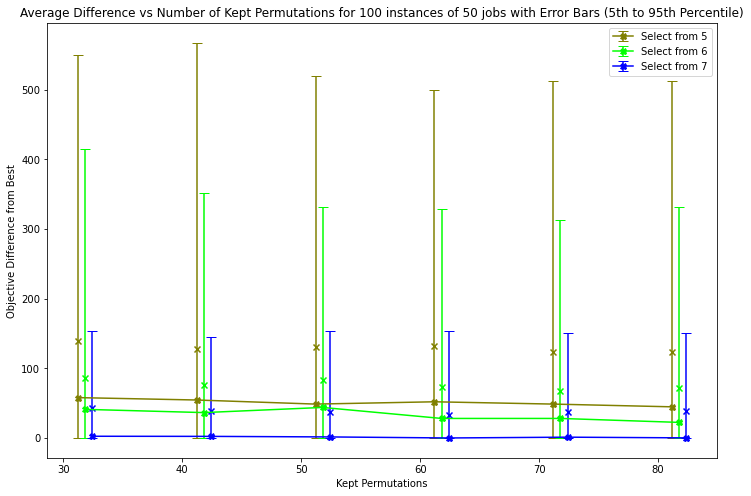}
	\vspace{-3mm}
	\caption{Comparison of 5th to 95th percentiles of objective minus best values for the selection algorithm run with different parameter values (number of kept permutations $K$ and permutation window $W$) for 100 50-job scenarios. Best overall values were taken as the lowest objective function obtained from all insertion and selection algorithm variants for the corresponding 50-job scenario. Medians are plotted with dots and connected by lines, and means are plotted with $\times$'s. Increasing $W$ produces notable reductions in error, while increasing $K$ above 30 has little effect.}
	\label{fig:AvgObjDif_NKEEP_Select}
\end{figure}

\begin{figure}[ht]
	\centering	\includegraphics[width=1.0\linewidth]{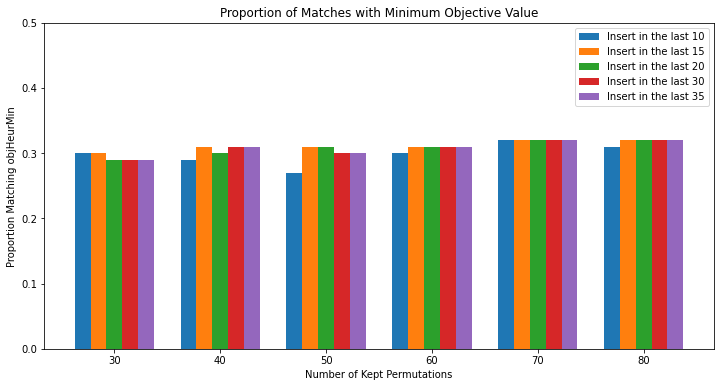}
	\vspace{-3mm}
	\caption{Proportion that each insertion algorithm attains the minimum penalty out of all obtained values across 50 job instances. The proportion evens out at around 0.3.}
	\label{fig:Prop_Insert}
\end{figure}

\begin{figure}[ht]
	\centering	\includegraphics[width=1.0\linewidth]{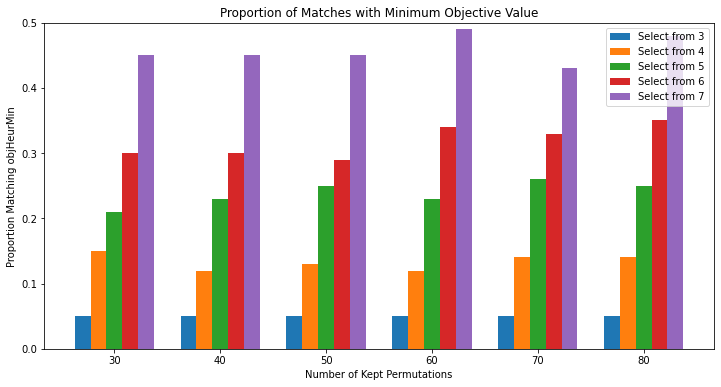}
	\vspace{-3mm}
	\caption{Proportion that each selection algorithm variant attains the minimum penalty out of all obtained values across 50 job instances. Selection continues to improve with larger windows, and reaches almost 0.5 with $W=7$.}
	\label{fig:Prop_Select}
\end{figure}

Figures \ref{fig:AvgObjDif_NKEEP_Insert} through \ref{fig:Pair_I15_S7_Keep-50-60-70} show performance comparisons between insertion and selection heuristic algorithms with different parameter values, for 100 instances of 50 job scenarios generated using the distributions described in Table~\ref{tab:1}. 
The same $100$ instances were run using all algorithm variants, and the best prediction from all variants was stored.  

Figure \ref{fig:AvgObjDif_NKEEP_Insert} shows the difference between the objective values for several insertion algorithm variants with different values of $P$ and $S$, compared to the best objective obtained for each instance over all insertion and selection algorithms. The figure shows mean and median differences for each variant, as well as the 5th and 95th percentile values.
Maximum number of kept permutations varies from 30 to 80, while maximum number of insertions varies from 10 to 35. Altogether 30 combinations of parameter values are used.
Results show little change in all performance measures when 50 or more permutations are kept ($P \ge 50)$, and when insertions are limited to the last 15 or more slots ($S \ge 15$). The mean values for the different algorithm variants are consistently higher than the medians, indicated by right-skewed distributions.

Figure \ref{fig:AvgObjDif_NKEEP_Select} parallels Figure \ref{fig:AvgObjDif_NKEEP_Insert} by showing means, medians, and 5th and 95th percentile values for objective function differences from best obtained using the selection algorithm variants with different values of $K$ and $W$. The selection algorithm with $W=7$  is far better than $W \le 6$. Indeed, comparison with Figure \ref{fig:AvgObjDif_NKEEP_Insert} shows that $W=7$ also beats all insertion variants.  The number of kept permutations $K$ has little effect on algorithm performance, regardless of the selection parameter. 

%Comparisons are also run between the selection and insertion algorithms with $100$ instances of $50$ jobs. The insertion algorithm was run trying the last $10$, $15$, $20$, $30$, and $35$ possible insertions and the selection algorithm was run with a selection window of $3$, $4$, $5$, $6$, and $7$. All of these variations were run with six different values of permutations to keep ($30$, $40$, $50$, $60$, $70$, and $80$). Using the same $100$ instances, the best prediction from all the different runs on each instance was saved as the best solution and used for analysis. Most figures for selection are focused on windows of $5$, $6$, and $7$ due to the scale being much larger for windows of $3$ and $4$. Figures 

Figure \ref{fig:Prop_Insert} shows the proportion of times each insertion algorithm variant matched the overall best prediction. Each variant matches the best obtained objective value for about 30 percent of instances regardless of the values of $P$ and $S$. 
We do see a slight dip in performance for $S=10$ compared to $S>10$, and an even smaller dip in performane when $P<70$. This is consistent with what was seen in Figure~\ref{fig:AvgObjDif_NKEEP_Insert}, which shows that the distribution of errors for the insertion algorithm is optimized for $S\ge 10$ and $P\ge 60$.

On the other hand, Figure~\ref{fig:Prop_Select} shows that big improvements in matching performance are obtained when the select from parameter is increased.  When the select from parameter is 7, the overall best performance is matched at least 45 percent of the time across the range of kept permutation values. Selecting from $6$ has similar performance to the insertion algorithms.

\subsubsection{Pairwise comparisons of algorithm performance}

\begin{figure}[ht]
	\centering	\includegraphics[width=1.2\linewidth]{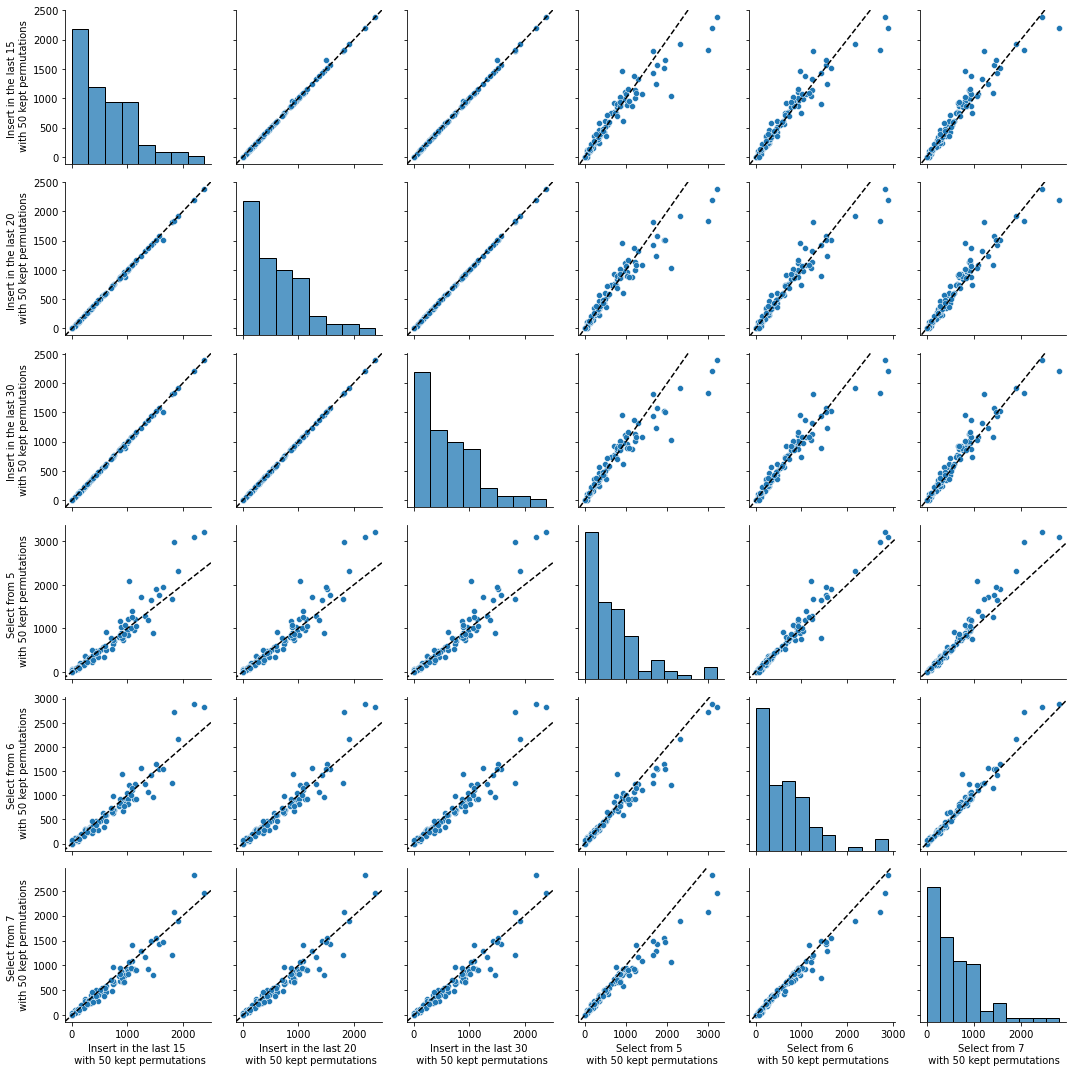}
	\vspace{-1mm}
	\caption{Scatter plots comparing  penalties obtained from different insertion and selection algorithms. Pair plots for insertion algorithm variants with $S=15,20,30$ and $P=50$ and selection algorithm variants with $W = 5,6,7$ and $K = 50$. }
	\label{fig:Pair_I15-20-30_S5-6-7_Keep50}
\end{figure}

\begin{figure}[ht]
	\centering	\includegraphics[width=1.2\linewidth]{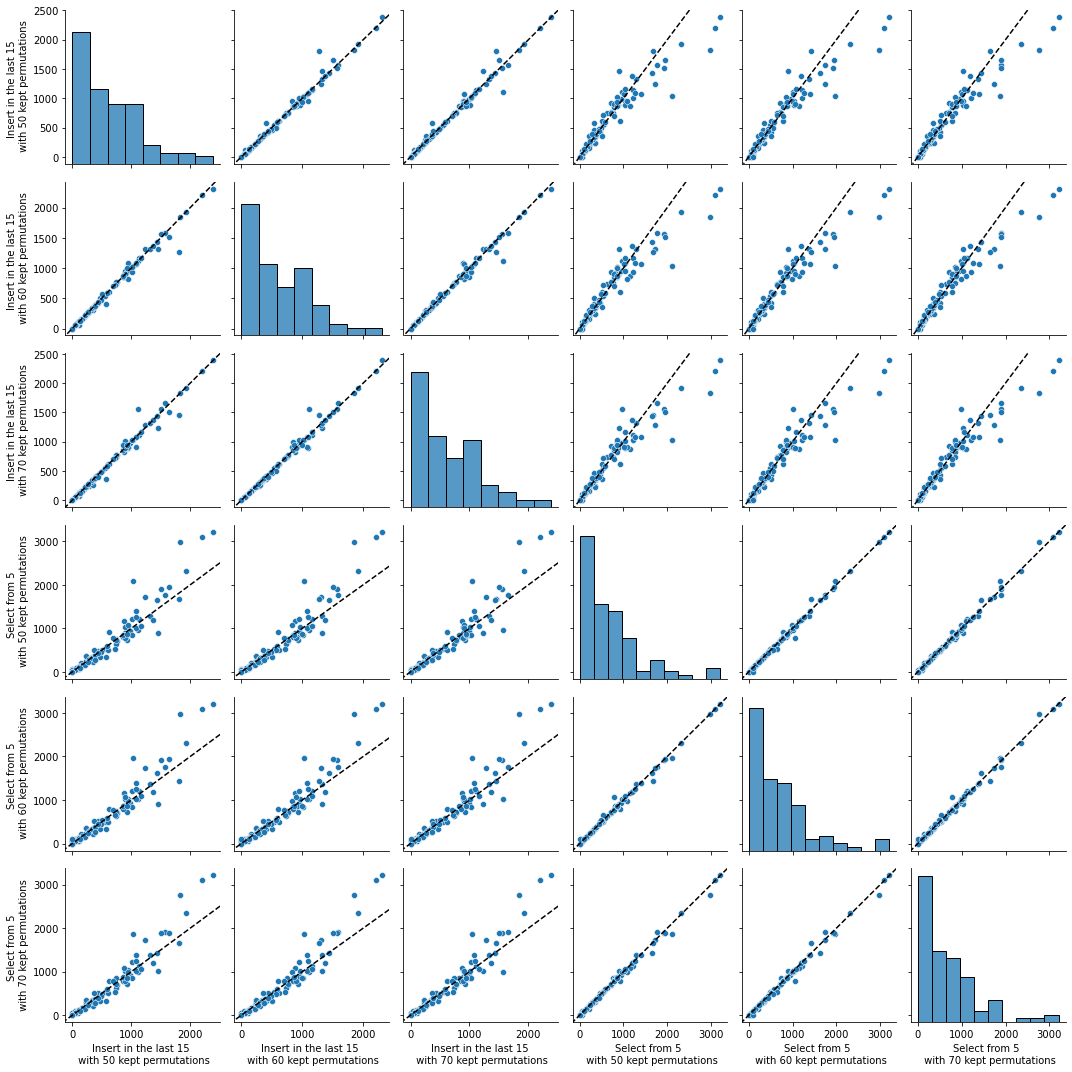}
	\vspace{-3mm}
	\caption{Scatter plots comparing  penalties obtained from different insertion and selection algorithms. Pair plots for insertion algorithm variants with $S=15$ and $P=50,60,70$ and selection algorithm variants with $W = 5$ and $K = 50,60,70$. } 
	\label{fig:Pair_I15_S5_Keep-50-60-70}
\end{figure}

\begin{figure}[ht]
	\centering	\includegraphics[width=1.2\linewidth]{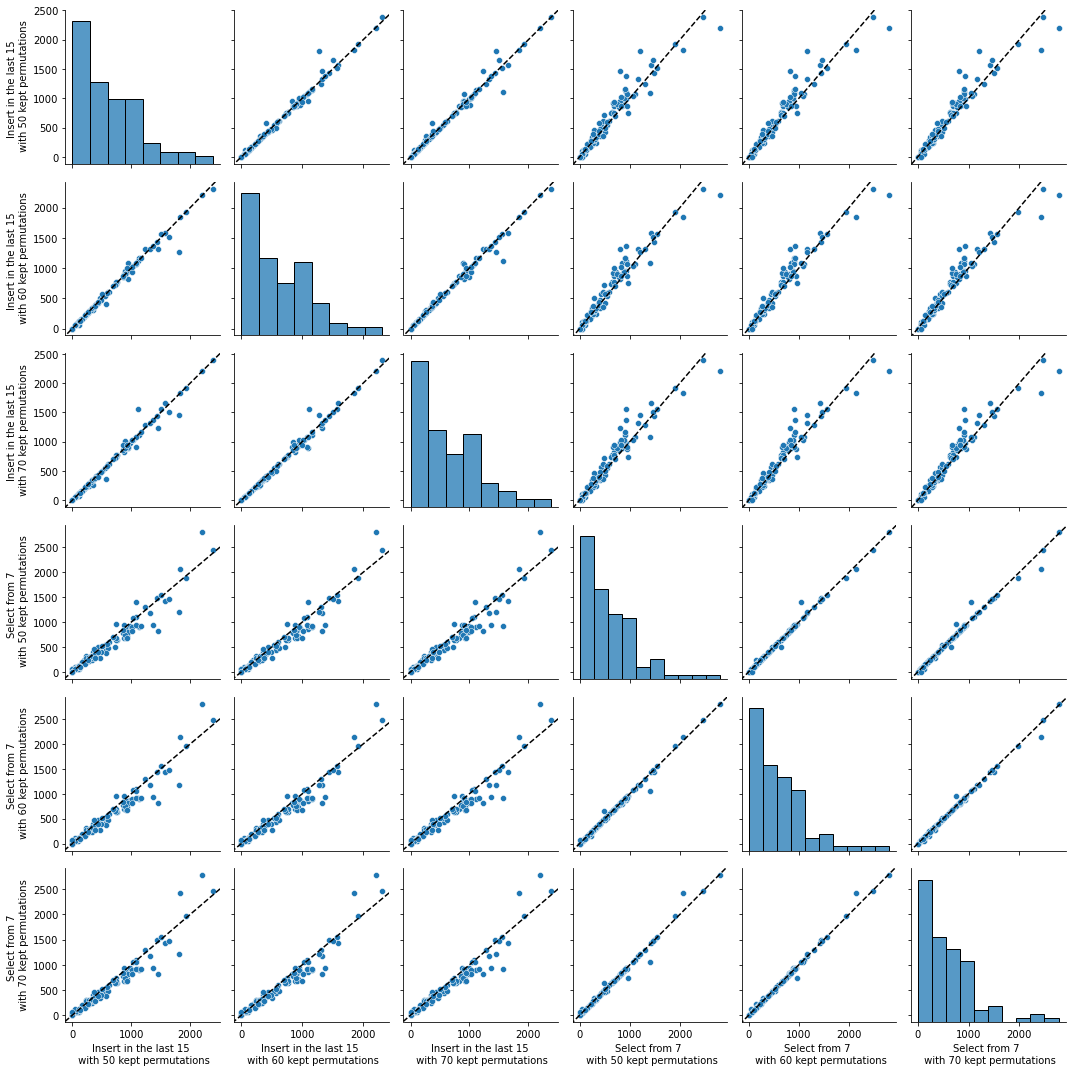}
	\vspace{-3mm}
	\caption{Scatter plots comparing  penalties obtained from different insertion and selection algorithms. Pair plots for insertion algorithm variants with $S=15$ and $P=50,60,70$ and selection algorithm variants with $W=7$ and $K=50,60,70$. }
	\label{fig:Pair_I15_S7_Keep-50-60-70}
\end{figure}

Figures \ref{fig:Pair_I15-20-30_S5-6-7_Keep50} through 
\ref{fig:Pair_I15_S7_Keep-50-60-70}
show scatter plots that provide detailed comparisons between pairs of algorithm variants. Each point in the scatter plot represents one particular scheduling scenario . The $x$ coordinatel of the point represents the penalty achieved by one algorithm variant, and the $y$ coordinate of the point shows the penalty achieved on the same scenario by a different algorithm variant.

Figure \ref{fig:Pair_I15-20-30_S5-6-7_Keep50}
compares insertion variants with  $S=15,20,30$ compared to the selection variants with $W=5,6,7$ with $P=K=50$.  The 3 variations of the insertion heuristic give almost identical results (since points align closely with the 45 degree line), while the selection algorithm improves significantly  as $W$ increases. Selection with a window of $7$ performs better than the insertion variations, but insertion is better than the other two selection variations.

Figure \ref{fig:Pair_I15_S5_Keep-50-60-70} compares insertion variants with  $S=15$ compared to the selection variants with $W=5$ with $P=50,60,70$ and $K=50,60,70$.  The 3 variations of the insertion heuristic give almost identical results (since points align closely with the 45 degree line), while the selection algorithm improves significantly  as $W$ increases. Selection with a window of $7$ performs better than the insertion variations, but insertion is better than the other two selection variations. While there is little variation over the kept permutations for either algorithm, inserting in the last $15$ is consistently better than  selection with a window of $5$, especially with larger objective values.

Figure \ref{fig:Pair_I15_S7_Keep-50-60-70} compares insertion variants with  $S=15$ compared to the selection variants with $W=7$ with $P=50,60,70$ and $K=50,60,70$.  The 3 variations of the insertion heuristic give almost identical results (since points align closely with the 45 degree line), while the selection algorithm improves significantly  as $W$ increases. Selection with a window of $7$ performs better than the insertion variations, but insertion is better than the other two selection variations. 
 While there is little variation over the kept permutations for either algorithm, inserting in the last $15$ is consistently worse than the selection algorithm with a window of $7$, especially with smaller objective values.

\subsubsection{Runtime comparisons for insertion and selection}

\begin{figure}[ht]
	\centering	\includegraphics[width=1.0\linewidth]{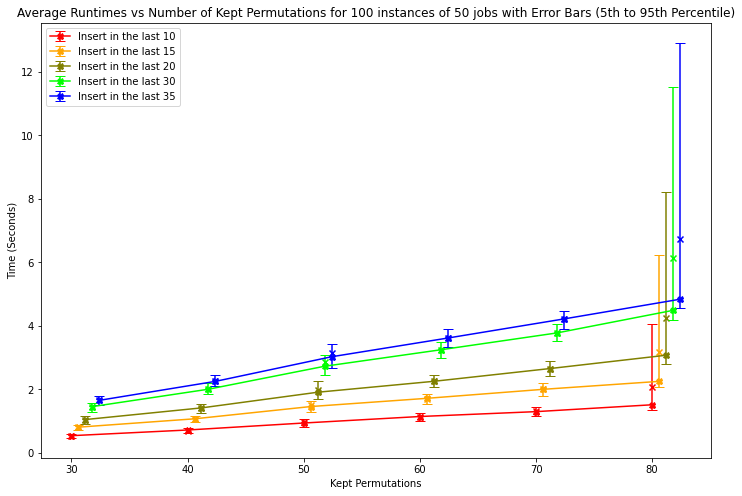}
	\vspace{-3mm}
	\caption{Runtime comparisons for insertion algorithm variants with kept permutations $P= 30,\ldots 80$  and  $S = 10,\ldots 35$. The median runtimes for each variant are plotted with circles, means are plotted with $\times$'s. and error bars indicate the 5th and 95th percentiles. The runtime is linear in $P$ for fixed $S$, with slope that increases as $S$ increases.}
	\label{fig:runtime_NKEEP_Insert}
\end{figure}

Figure \ref{fig:runtime_NKEEP_Insert} shows runtimes for the insertion algorithm using different values for the number of kept permutations $P$ and number of insertion slots $S$. The runtime is linear $P$ for fixed $S$, with slope increasing with $S$.  It appears that runtime is also linear in $S$ for fixed $P$. This linearity in parameters implies that implementations with large values of $P$ and $S$ are feasible. However, Figure~\ref{fig:AvgObjDif_NKEEP_Insert} indicates that large values are not necessary for good performance.

\begin{figure}[ht]
	\centering	\includegraphics[width=1.0\linewidth]{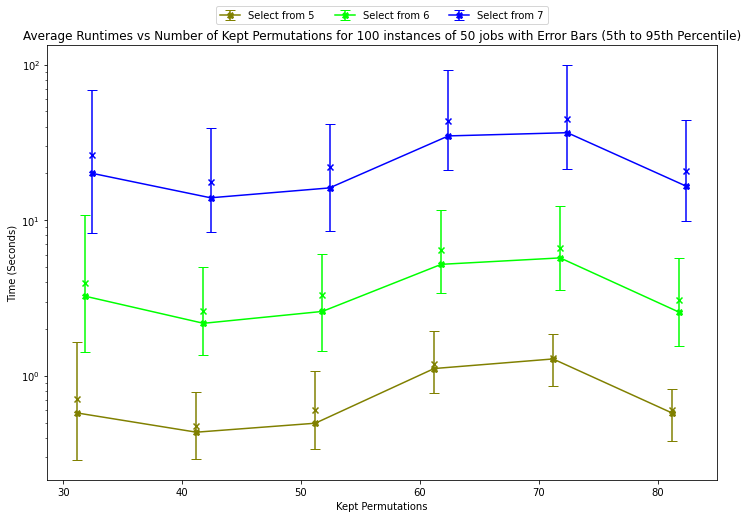}
	\vspace{-3mm}
	\caption{Runtime comparisons for selection algorithm variants with kept permutations $K= 30,\ldots 80$  and  permutation window size $W = 5,6,7$. The median runtimes for each variant are plotted with circles, means are plotted with $\times$'s, and error bars indicate the 5th and 95th percentiles. The runtime is to be mostly unaffected by the number of kept permutations, but increases exponentially with the window size.}
	\label{fig:runtime_NKEEP_Select}
\end{figure}

Figure \ref{fig:runtime_NKEEP_Select} shows the runtime for the selection algorithm with different values for permutation window $W$ and number of kept permutations $K$. The runtime is relatively constant for $K\ge 30$, wheras the runtime increases exponentially with $W$. Together with Figure~\ref{fig:AvgObjDif_NKEEP_Select}, these results indicate that improvements in selection algorithm performance can be acheived by increasing $W$, but at relatively high computation cost.

\section{Conclusions}
We have shown that the insertion and selection algorithms perform nearly optimally when the number of jobs is small, and offer a substantial improvement over dispatch rules as the number of jobs increases.  For both algorithms, parameters can be adjusted to effect tradeoffs between execution time and objective function performance, although with insertion the performance improvements that can be achieved by parameter adjustment are limited.  In head-to-head comparisons between insertion and selection, selection and insertion algorithms have comparable accuracy when comparing variants with comparable runtime.  For example, The insertion algorithm with $P=50$ and $S=15$ has comparable performance and runtime with the selection algorithm with $W=6$ and $K=30$.  However, the performance of the insertion algorithm is limited and cannot be improved by further increasing the selection window and kept permutations. On the other hand, the selection algorithm can be further improved by increasing the permutation window--but at high cost in execution time. 

Although we did not do an extensive study of possible neural network architectures, the results from our simple architecture suggest that neural networks may not be well suited to this application.
Preliminary investigations with much larger networks also showed very poor performance.
It is reasonable that neural networks would have difficulty solving scheduling problems, due to their discontinuous nature.  Neural networks training algorithms presuppose a continuous dependence of the  loss function on inputs, so that gradient descent can be used to obtain incremental improvements.  However, in scheduling problems the penalty is a discontinuous function of the inputs, so gradient-based parameter optimization cannot be expected to work.

Projected future work includes more extensive benchmarking.  In the literature, other methods are suggested to generate scenarios \cite{fernandez2015neh}, which may be used to test the algorithms over a wider variety of conditions.

The algorithms may also be further optimized. Further optimization can target reducing execution time and/or improving performance.  
\begin{itemize}
\item
As far as reducing execution time, not all of the speedup features for insertion and selection described in Section~\ref{sec:algDescriptions} have been implemented. There are other possibilities for speedup as well.  For instance,  in the selection algorithm there may be a way of selectively generating permutations,  rather than generating all permutations in the permutation window.  
\item
Performance may perhaps be improved by combining the two algorithms. It is possible, for instance, that applying selection on the schedule generated by insertion may have the effect of "fine tuning", since selection performs a local reshuffling. 
\end{itemize}
Finally, the two algorithms may be generalized and applied to other problems, such as the permutation flowshop problem.  This research is currently in progress.

\section{DECLARATIONS}
\subsection*{Ethics approval and consent to participate}
Not applicable
\subsection*{Consent for publication}
Not applicable
\subsection*{Availability of data and materials} 
The code used in this paper to generate data and to apply and compare scheduling algorithms is stored in GitHub at:  \url{https://github.com/christhron/Flowshop_Iterative_Heuristics}.
\subsection*{Competing interests}
The authors declare no competing interests.
\subsection*{Funding}
No funding was received for conducting this study.
\subsection*{Authors' contributions}
\noindent
\textbf{Conceptualization}: OBO, CT,MA; 
\textbf{Formal analysis}: MG, GS, RW, CT;
\textbf{Investigation}: MG, GS, CT;
\textbf{Methodology}: MG, GS, CT;
\textbf{Software}: MG, GS,CT; 
\textbf{Supervision}: MA, CT; 
\textbf{Validation}: MG, GS, CT;
\textbf{Visualization}: MG, GS,CT; \textbf{Writing –
original draft}: MG, OBO, RW, CT; \textbf{Writing – review \& editing}: MG,CT,MA.
\subsection*{Acknowledgements}
We wish to thank the anonymous reviewers for valuable suggestions that greatly improved the clarity of the paper.

\section{Figures}

\clearpage
\bibliographystyle{unsrt}
\bibliography{citations}

\end{document}